# Evaluating Jacob Bernoulli's Ship Propulsion Artifice: Euler's Analysis and Critique

Sylvio R. Bistafa

sylvio.bistafa@gmail.com

## Abstract

The present work examines and compares the approaches of Jacob Bernoulli and Leonhard Euler to the problem of ship propulsion generated by internal forces. Jacob Bernoulli's analysis, developed in the late 17th century, relies on geometric interpretations and algebraic relationships to estimate the impulse exerted by a pendulum within a ship. His results, however, are shown to overestimate the force by scaling it with the height of fall rather than the velocity at the end of the fall and by amplifying it with a time dependence factor that proves to be erroneous. Euler's subsequent analysis, conducted nearly 50 years later, provides a more rigorous and mechanically sound treatment of the problem. Utilizing differential equations and a deeper understanding of the principles of motion, Euler demonstrates that no net motion can be generated by internal forces alone, thus refuting the concept of perpetual motion as proposed by Jacob Bernoulli. By comparing the two methodologies, this work highlights the evolution of mathematical and physical analysis from the intuitive but flawed approaches of the early calculus pioneers to the more precise and reliable methods established by later mathematicians. Euler's work not only corrects the errors in Jacob Bernoulli's analysis but also solidifies the understanding that propulsion requires external forces, a conclusion consistent with the laws of Newtonian mechanics.

## 1. Introduction

Jacob Bernoulli (1655–1705) is widely regarded as one of the foremost mathematicians of the 17th century, standing alongside luminaries such as Newton, Huygens, Leibniz, and his own younger brother Johann, whom he tutored in mathematics. He was a strong advocate of Leibnizian calculus during the Leibniz–Newton calculus controversy and is considered one of the founders of the calculus of variations. Jacob Bernoulli's mathematical work often followed the geometric constructions pioneered by Descartes, who introduced this problem-solving technique in *La Géométrie* (1637). Utilizing these constructions, Jacob Bernoulli frequently employed infinite series solutions for the resulting integrals in terms of the known functions of his time, such as sine, exponential, and inverse sine functions. His work on the *Elastica* is notable as one of his attempts to address more physically grounded problems. In this complex development, which extensively uses geometric constructions and mechanical concepts like levers and pulleys, Jacob Bernoulli was the first to discover that the solution to the *Elastica* could be expressed using elliptic integrals.

Despite his significant achievements in mathematics, Jacob Bernoulli's contributions to applied topics are less distinguished. He seemed particularly intrigued by the idea of perpetual motion devices. For instance, in *Acta Eruditorum*[a] Jacob Bernoulli critically examined a perpetual motion device composed of a bellows, a lever with an attached weight, and a mercury container connected to the bellows via tubing (Fig. 1). Although the device's flaw had already been pointed out earlier, Jacob Bernoulli's examination was meticulous, focusing on the forces resulting from unequal pressure distribution on the bellows' surfaces. He provided a solution using one of his preferred mathematical tools—an infinite series in arithmetic progression. Despite offering several criticisms, Jacob Bernoulli did not outright dismiss the idea. In fact, by suggesting improvements to the device, he seemed to tacitly admit that it might work. As is well known, perpetual motion machines are impossible according to the laws of physics, yet such devices were often devised in the past—and still are today—to create an illusion of continuous motion.

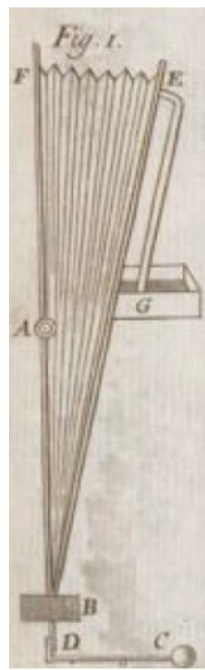

Figure 1: A perpetual motion device, composed of a bellows, a lever with an attached weight, and a mercury container connected to the bellows via tubing.

The present work will examine another of Jacob Bernoulli's proposals, based on a different physical misconception—that a device could be arranged to propel a ship using only internal motion. In Article XXVII of *Varia Posthuma*[b], Jacob Bernoulli presents an analysis of a device designed to propel a ship through motion confined entirely within the vessel itself (Fig. 2). This concept, developed in the late 17th and early 18th centuries, reflects Jacob Bernoulli's broader interest in mechanics and mechanisms. His analysis explores the feasibility of generating motion without relying on external forces like wind or oars—a topic of considerable curiosity during that era.

---

[a] *Examen Bernoullianum. Acta Eruditorum, Leipzig, 1686 Dec.*, p. 625. Available at: https://archive.org/download/s1id13206520/s1id13206520.pdf

[b] *Artificium impellendi Navem a principio motus intra ipsam Navem concluso* (A device for propelling a ship by internal motion confined within the ship itself.), *Articulus XXVII*, *Varia Posthuma Jacobi Bernoulli*, in *Jacobi Bernoulli Opera Tomus Secundus*, Publisher: Cramer & Philibert (*Genevae*), 1744, pp. 1109 – 1115. A posthumous collection of works by the Swiss mathematician Jacob Bernoulli (1655–1705). Available at:

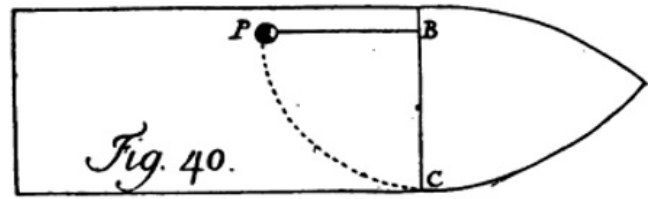

Figure 2: Original sketch of the device attached to the ship by Jacob Bernoulli.

Euler's involvement with this device is documented in a 1750 publication—*Examen artificii navis a principio motus interno propellendi…*[c] Skeptical about the feasibility of such a mechanism, Euler undertook a closer examination to evaluate it according to the laws of motion. In §.9, Euler explains that "…this suspicion is further heightened by the fact that the description of this device is only found in the posthumous works of Jacob Bernoulli and was never published during his lifetime. It seems highly improbable that such a great man, now deceased, would have concealed such an invention—one that would certainly surpass all of his other inventions, even the greatest—unless he himself had doubts about its success."

Euler also notes in §.18 that "…Cramer, the commentator of Jacob Bernoulli's posthumous work, clearly noticed an error in determining the propelling force arising from the impact but did not correct it due to the difficulty of the necessary calculations." Euler emphasizes that disproving the mechanism would not diminish Jacob Bernoulli's reputation, as he was cautious about releasing unproven ideas.

Euler belonged to the second generation of the early pioneers of modern science. He is widely regarded as one of the greatest mathematicians of all time, with his influence extending well beyond mathematics into general physics during the 18th century. Euler made remarkable contributions to fields such as mechanics in general, celestial mechanics, fluid dynamics, acoustics, and optics. In addition to his theoretical work, Euler was a practical and down-to-earth scientist, demonstrating his engineering prowess through contributions to the theory of machines and naval science. He was also deeply concerned with addressing the practical challenges of daily life in his era, such as developing methods for raising and distributing water for

---

https://gdz.sub.uni-goettingen.de/id/PPN585140006
https://archive.org/details/bub_gb_VNsVDxsTdBgC/page/n503/mode/2up?view=theater

[c] E137-- *Examen artificii navis a principio motus interno propellendi quod quondam ab acutissimo viro Iacobo Bernoulli est propositum* (Examination of an artifice for propelling a ship by the principle of internal motion which was once proposed by the most astute man, Jacob Bernoulli". In *Novi Commentarii academiae scientiarum Petropolitanae* 1, 1750, pp. 106-123.

agricultural and public consumption. Through these achievements, Euler not only validated and consolidated the discoveries of earlier pioneering scientists but also established methods and approaches that remain foundational in their respective fields to this day.

Euler's involvement with Jacob Bernoulli's ship propulsion device is well justified by his polymathic nature and his interest in applying scientific principles to real-world problems. In the following section, we will analyze the main points of Jacob Bernoulli's proposal, contrasting it with Euler's critique, made about 50 years after Jacob Bernoulli's death.

## 2. Comparative Analysis

Utilizing an early form of calculus based on geometric interpretations of the phenomena involved, Jacob Bernoulli's analysis becomes intricate and, at times, cumbersome for the modern reader. This analysis is accompanied by Cramer's commentary. Following this, Euler's analysis, as presented in *Examen artificii navis a principio motus interno propellendi*, is examined to compare the differing methodologies in addressing the same problem.

### 2.1 Jacob Bernoulli's analysis

**The impulse imparted to the ship by the pendulum.**

Let the height of the pendulum's perpendicular descent, or its length $BC = a$, be represented by the triangle $KLM$ (Fig. 3), and the time of the pendulum's perpendicular descent through $BC$ be represented by $KL = t$. The velocity at the end of the fall is $LM = 2a/t$; and from this, the time of descent and ascent of the pendulum through the quadrant, or rather the complete oscillation, be denoted by $T$. The impulse imparted to the ship by the pendulum in the direction of the bow is $2aT/t$.

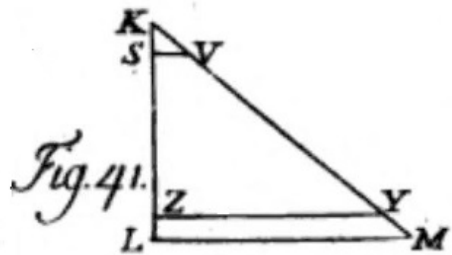

Figure 3: Jacob Bernoulli's auxiliary sketch to model the impulse imparted to the ship by the pendulum.

**Cramer's comments:**

(a) It is known from Galileo's theory that, under the hypothesis of constant gravity, if the times of perpendicular descents are represented by the segments KL, KZ, KS, the velocities acquired can be represented by the perpendiculars of some triangle LM, ZY, SV, and the distances covered by the areas of the triangles KLM, KZY, KSV.

(b) By the impulse towards the bow of the ship, which is represented as $2aT/t$, is meant the distance that the ship, moving with a velocity $2aT/t$, would cover in time $T$ in a non-resisting medium, just as shortly after, by the force of gravity, is meant the small distance that a heavy body, descending vertically, would cover in the first small time $dt$, which is represented by the triangle KSV.

**Comments:**

Velocity at the end of a fall $v$, given as $v = 2a/t$, can be explained by relating $v$ to the average velocity $v_{avg}$ during the fall, which is given by $v_{avg} = a/t$. Since $v = 2 \times v_{avg}$, then $v = 2a/t$.

The pendulum's impulse, expressed as $2aT/t$, appears somewhat speculative since Jacob Bernoulli did not provide a detailed explanation for this result. It's challenging to discern exactly what reasoning led him to this conclusion. This lack of elaboration is characteristic of Jacob Bernoulli's approach in several of his works, where he often presented results without thoroughly documenting the underlying thought process.

- The energy available in the system is related to the period $T$, as the potential energy at the highest point is converted into kinetic energy at the lowest point. The time $t$ affects how rapidly this energy is transferred.
- $T$ reflects the full oscillation cycle, which is important for understanding the complete behavior of the pendulum over time.
- The impulse is not only about the force at a single moment but about how that force accumulates over time. The period $T$ provides a natural scaling factor when considering how the force is distributed over the entire oscillation.
- The formula $2aT/t$ implies that the impulse involves both the specific time of force application (descent time $t$) and the overall cycle of motion (period $T$).
- Since the velocity at the lowest point of the swing depends on $t$ (related to the height $a$), and $T$ dictates how the pendulum returns, both are integral to describe the pendulum's motion and the resulting impulse.

If we simply omit the time dependence factor $T/t$ from $2aT/t$, the impulse would simplify to $2a$. This approximation would be nearly correct if $\sqrt{a}$ (the velocity at the end of the fall) were used instead of the height of fall $a$ in the final result. The velocity $\sqrt{a}$ accurately reflects the energy conversion from potential to kinetic energy as the pendulum strikes the platform, capturing the true dynamics of the system. By considering $\sqrt{a}$ rather than $a$, the analysis would align more closely with the principles of energy conservation and momentum, leading to a more physically accurate representation of the impulse.

**The opposing impulse towards the stern, produced by the force of gravity.**

The development begins by defining a small time interval $dt$ and relates it to the gravitational force acting on a ship. The motion is then decomposed into components, focusing on how these components resolve along different paths (e.g., PD & PF, and further into PG & PH), with various geometric and algebraic relationships between these segments being established (Fig. 4). Velocities at different points, such as P and C, are compared, with the velocity at C derived from the height of fall. The distance traveled during the small interval $dt$ is calculated directly and through other motion components. Using these relationships, the impulse exerted by gravity on the ship during dt is determined. Finally, by integrating the total impulse over the motion (covering both descent and ascent through a quadrant), it is concluded that the total impulse due to gravity over the complete oscillation period $T$ is proportional to $\frac{2}{3}a$, where $a$ represents the height of fall. This comprehensive approach illustrates the classical methods of motion decomposition, velocity and distance calculations, impulse determination, and integration of the total effect.

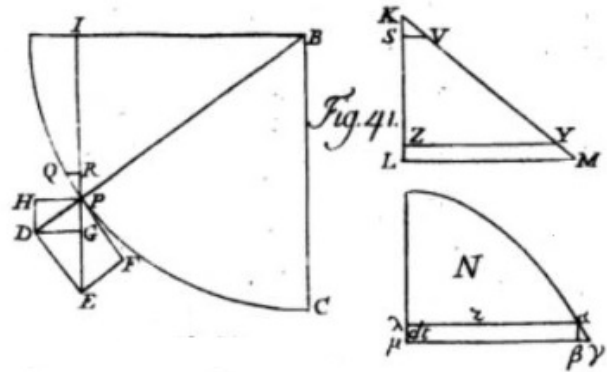

Figure 4: Jacob Bernoulli's auxiliary sketch to model the impulse toward the stern by the force of gravity on the pendulum.

**Cramer's comment:**

(c) Since the effects of all preceding impulses are understood to be represented by the curved area *N*, or the space covered by the ship moving towards the stern during the entire time of these impulses, it is evident that if no new impulse were to occur in the last moment *dt*, the ship would continue to move at the same speed *z*, which it had just acquired, and during this last moment, it would cover the space represented by the parallelogram. But because, due to the new impulse, the curved area *N* increases not only by the parallelogram $\lambda\mu\beta\alpha$ but also by the trilinear area $\alpha\beta\gamma$, this trilinear area must therefore be considered the effect of the last impulse.

**Comment:**

The conclusion that the gravitational force over the complete oscillation period $T$ is proportional to $\frac{2}{3}a$ suggests that this force depends directly on the height of the fall, $a$. However, this interpretation oversimplifies the dynamics involved. In reality, the force should be proportional to the velocity at the end of the fall, which is $\sqrt{a}$. The velocity $\sqrt{a}$ better captures the energy conversion from potential to kinetic energy as the pendulum swings, reflecting the actual force applied as the pendulum reaches its lowest point. Thus, the result $\frac{2}{3}a$ might overestimate the force by focusing on height rather than the velocity that height generates, leading to an overestimation of the impact on the system.

**The opposing impulse towards the stern, produced by the centrifugal force.**

Let QD be the tangent to the circle (Fig. 5), and thus PD represents the centrifugal effort, which is further resolved into two other efforts PG and PH; of these, PH is engaged in pulling the ship towards the stern, while PG is used in pressing the ship downward against the water. As is known, $PD = PQ^2 : 2BP = ds^2 : 2a$; from which $BP\,[a] : BI[x] = PD\,[\frac{ds^2}{2a}] : [PH]\,[\frac{xds^2}{2a^2}]$ = [because $ds^2 = 2dt\sqrt{ay} : t$], this will be equal to $dsdt\sqrt{ay} : a^2 t$; which is precisely double the other impact PH, caused by the force of gravity; hence the total impulse of the centrifugal effort exerted over time $T$ is double the total impulse arising from gravity, and thus the sum of both $\frac{2}{3}a + \frac{4}{3}a = 2a$.

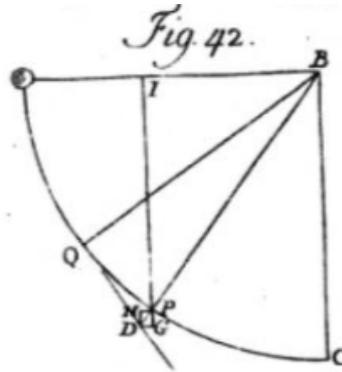

Figure 5: Jacob Bernoulli's auxiliary sketch to model the impulse toward the stern by the centrifugal force.

**Comment:**

The conclusion that the centrifugal force is twice the gravitational force is correct. However, the analysis is compromised by its incorrect dependence on $a$ rather than $\sqrt{a}$. The force should be proportional to the velocity, which is related to $\sqrt{a}$, rather than directly proportional to the height $a$. This miscalculation leads to an overestimation of the force, as the correct physical interpretation should consider the energy dynamics of the system, where the velocity at the end of the fall (proportional to $\sqrt{a}$) plays a crucial role in determining the force rather than just the height.

**The resulting force acting on the ship.**

Therefore, if you subtract these impulses, which propel the ship towards the stern over time $T$, from the one that propels it towards the bow during the same time, which is found to be $2aT/t$, the effective part of this impulse towards the bow remains as $2aT/t - 2a = 2a(T-t)/t$.

From this, it is evident that an impulse can be generated in the direction of the ship's bow because $T$ is greater than $t$; and this impulse can be defined as long as the ratio between $T$ and t is known.

Since the small time interval $dt$, during which the arc of the quadrant QP or $ds$ is described, is found to be $tds : 2\sqrt{ay}$ = [from the nature of the circle] $ady : \sqrt{a^2 - y^2}$, it follows that $dt = atdy : 2\sqrt{a^3 y - ay^3}$ = [given $ay = u^2$] $atdu : \sqrt{a^4 - u^4}$, which corresponds to the element of the curve Elastic curve, whose integral [according to Prop. LVII of the Infin. Series] is approximately $\frac{131}{100}t$ and hence the double of this, namely the total time for the quadrant to be traversed twice by ascending and descending, is $T = \frac{262}{100}t$, and so finally, the quantity of impulse of the pendulum towards the bow, which is imparted during each period $T$, is $2a(T-t)/t = \frac{324}{100}a =$ [if force] $c$.

**Comments:**

Jacob Bernoulli's analysis calculates the striking impulse of the pendulum as $2aT/t$. However, if we assume small oscillations (which is not the case here), where $T = 2t$, the impulse would simplify to $4 \times a$.

On the other hand, Jacob Bernoulli's time dependence factor, given as $T = 2.62 \times t$, taken from his work with the *Elastica*, where the solution is given in terms of elliptic integrals. This yields an impulse of $5.24 \times a$. However, using the modern correction factor, where $K(k)$ is the complete elliptic integral of the first kind (and for an amplitude $\theta_0 = \frac{\pi}{2}$, $sin\frac{\pi}{4} = \frac{1}{\sqrt{2}}$), the elliptic integral is approximately 3.708. This gives $T \approx 3.708 \times t$, and the impulse as $7,416 \times a$.

It is therefore evident that Jacob Bernoulli's inclusion of the time dependence factor $T/t$ in his analysis significantly overestimates the pendulum striking impulse. As discussed earlier, if this time factor were omitted, the resulting impulse would be $2a$ as expected (or $2\sqrt{a}$ if the correct scaling based on velocity, not height, were applied).

The pendulum striking force of $2a$ would then be counteracted by the backward forces of gravity $\frac{2}{3}a$ and centrifugal force $\frac{4}{3}a$, leading to a net force of zero on the ship.

## 2.2 Euler's analysis

Figure 6 shows Euler's sketch of the machine proposed by Jacob Bernoulli, which was used in his analysis of the problem.

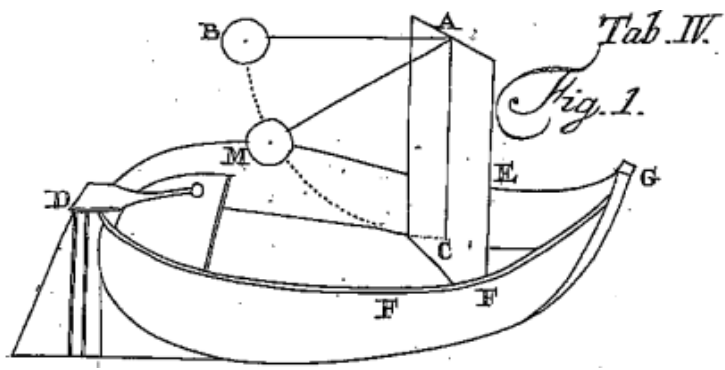

Figure 6: Sketch of the elements used by Euler in his analysis of the system for propelling ships by internal forces.

**The impulse imparted to the ship by the pendulum.**

Euler's analysis examines the force exerted by a pendulum on a ship when it strikes an immobile elastic platform. The pendulum, with mass M, impacts the platform with a velocity corresponding to the height a from which it fell. To explore the collision in more detail, Euler introduces an elastic element (Fig. 7) attached to the platform, which compresses upon impact. As the pendulum penetrates further, the force exerted by the elastic element, denoted as $P$, is related to the change in velocity and distance traveled during the collision.

Euler applies the laws of motion, establishing that $Mdv = -Pdx$, where $dv$ is the change in velocity and $dx$ is the small distance moved. The integral of the force over time $\int Pdt$ is calculated to determine the total impulse during the collision. Since the platform and pendulum are assumed to be perfectly elastic, the final result shows that after the collision, the pendulum's velocity changes direction but retains its magnitude. Euler concludes that the total instantaneous force propelling the ship due to the collision is $P = 4M\sqrt{a}$, where $\sqrt{a}$ is the initial velocity corresponding to the height of fall.

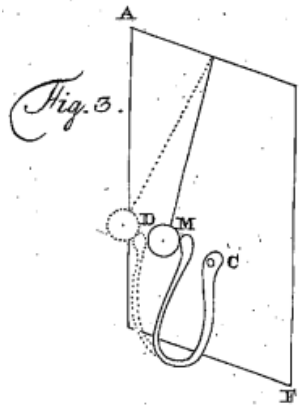

Figure 7: Elastic element CD is attached to the platform at point C.

**Impulse towards the stern produced by the gravitational force and the centrifugal force.**

1. **Gravitational force:**

As the pendulum moves, gravity acts downward with a force of *M* (Fig. 8). This gravitational force is resolved into two components: along the normal to the string (MQ): $M \sin \phi$, which accelerates the pendulum and does not contribute to the tension in the string. Along the string (AM): $M \cos \phi$, which contributes to the tension in the string.

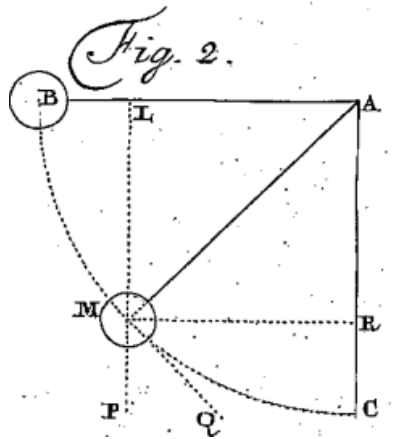

Figure 8: Euler's sketch for modeling the force of gravity and the centrifugal force acting on the ship.

2. **Centrifugal force:**

The velocity of the pendulum at point M corresponds to the height $M \cos \phi$, and the centrifugal force is given by $2M \cos \phi$, acting along the string AM.

3. **Total tension in the string:**

The total tension $T$ in the string AM is the sum of the forces along the string due to both gravity and the centrifugal force: $T = 3M \cos \phi$.

4. **The force acting on the ship, pulling it backward:**

Calculated by resolving the tension along the direction AM, giving: $F = 3M\cos \phi \sin \phi$. When this force is multiplied by the element of time $dt = -ad\phi\sqrt{a \cos \phi}$, this product gives the instantaneous backward force $-3Md\phi \sin \phi \sqrt{a \cos \phi}$. By substituting $\cos \phi = z$, the expression simplifies to $3Mdz\sqrt{az}$. Integrating this expression yields the total force pulling the ship backward as $2Mz\sqrt{az} = 2M \cos \phi \sqrt{a \cos \phi}$. At the lowest point of the pendulum's swing ($\phi = 0$), the total force from the complete descent is $2M\sqrt{a}$. Since the ascent contributes an equal force, the ship experiences a total backward force of $4M\sqrt{a}$ for each complete oscillation. If the ship were free to move, this force would impart a motion quantity of $4M\sqrt{a}$.

5. **Euler's conclusion:**

Euler then concludes that the motion generated by the pendulum's impact is countered by the forces during the pendulum's ascent and subsequent descent, resulting in no progressive movement of the ship. He also mentions an error Jacob Bernoulli made, which was identified by his commentator, Cramer.

## 2.3 Impulse imparted to the ship by the pendulum: Modern analysis

The **Dirac Delta function** $\delta(t)$ is a mathematical tool often used to represent an instantaneous impulse or force that acts over an infinitesimally short period of time. If the collision happens at time $t_0$, the force can be expressed as: $F(t) = P \cdot \delta(t -$ $t_0$), where $P$ is the magnitude of the force, $\delta(t - t_0)$, indicates that the force acts only at the instant $t = t_0$.

Impulse $I$ is defined as the integral of force over time. With the Dirac Delta function, the impulse delivered by the force $F(t)$ is:

$$I = \int_{-\infty}^{\infty} F(t)dt = \int_{-\infty}^{\infty} P \cdot \delta(t - t_0)dt.$$

Using the sifting property[d] of the Dirac Delta function, this simplifies to $I = P$. This means that the impulse $I$ is equal to the magnitude of the force $P$, implying that all the momentum change happens in an instant.

Consider a pendulum with mass $M$ falling from a height $a$. Just before impact, the pendulum reaches a velocity $v = \sqrt{2ga}$ due to gravitational acceleration $g$. The collision between the pendulum and the platform occurs over an extremely short time interval, which we denote as $\Delta t$. During this time, the force exerted by the pendulum on the platform spikes briefly and can be modeled as an instantaneous impulse.

From the momentum-impulse theorem $I = \Delta p = M \cdot \Delta v$. For an elastic collision where the velocity after the impact is $-v$ (reversing direction but maintaining magnitude), the change in momentum is: $\Delta p = M(v - (-v)) = 2Mv$. Since $I = 2Mv$, and from the earlier equation $I = P$, we equate:

$$P = 2M\sqrt{2ga} = 2Mv.$$

As we discussed earlier, Euler calculated the collision force as $4M\sqrt{a} = 4Mv$, which suggests a duplication of the expected result. This occurs because Euler typically expresses Newton's Second Law in the form $F = (\propto mdv)/dt$, where $\propto = 2$, given that $F$ represents the weight (not just the mass) of the body, and $v = \sqrt{x}$ is the velocity corresponding to the height $dx$ covered during the infinitesimal time $dt = dx/\sqrt{x}$. Therefore, if we were to adhere to the conventional form of Newton's Second Law, Euler's result would indeed be written as $2Mv$, aligning with the result obtained through modern analysis.

## 2.4 Comparative Analysis of Jacob Bernoulli's and Euler's Approaches to Pendulum-Induced Ship Propulsion

Jacob Bernoulli's analysis is developed using of archaic notation and is based on geometric interpretations by assuming that the pendulum's impulse can be described by algebraic relations. His analysis is very simplistic and highly intuitive, by simply calculating the downward swing impulse as proportional to the height of the pendulum's fall and relating this to the time of descent and the pendulum's period.

Euler's analysis, in contrast, is more sophisticated, shorter, and accurate. He models the pendulum's interaction with the ship using differential equations, accounting for the elasticity of the platform and the continuous force exerted during the collision. This gives a more realistic depiction of the forces involved.

---

[d] The sifting property states that if you have a function $f(t)$, and you integrate the product of $f(t)$ and $\delta(t - t_0)$ over the entire real line, the result is the value of the function $f(t)$ at the point $t = t_0$: $\int_{-\infty}^{\infty} f(t)\delta(t - t_0)dt = f(t_0)$.

It is shown that Jacob Bernoulli's analysis gives the pendulum's impulsive force as proportional to $\frac{2aT}{t}$, whereas Euler's analysis gives it as equal to $4M\sqrt{a}$, where $M$ is the mass of the pendulum, $a$ is the height of fall, $T$ is the pendulum's period and $t$ is the pendulum's time of descent. It is seen then that Jacob Bernoulli's analysis overestimates Euler's analysis by $\sqrt{a}$, that is, the impulsive force in Euler's result came out as proportional to the velocity with which the pendulum strikes the platform, whereas Jacob Bernoulli's analysis reveals it to be proportional to the square of this velocity. It was shown that another cause of the amplification of the pendulum's striking force is the inclusion of the of the time dependence factor $T/t$. Without considering both amplifications of the pendulum's striking force, Jacob Bernoulli's analysis would be consistent with Euler's.

Euler's method is more mechanically sound, as it considers the gradual compression of the elastic platform and the continuous interaction. By relating the force to the pendulum's velocity at impact, Euler's model gives a more accurate picture of how the forces behave over time, leading to a more realistic result.

However, the equality of Euler's impact force calculation and the result from the Dirac Delta function model suggests that the inclusion of elasticity in Euler's analysis was either superficial or improperly modeled. For elasticity to have a meaningful impact, it needs to be fully integrated into the model with appropriate parameters. The oversight of this integration highlights a potential limitation in Euler's approach and underscores the importance of thoroughly modeling all relevant physical properties when analyzing such problems.

Jacob Bernoulli concludes that the effective part of these impulses towards the bow remains as $\frac{324}{100}a$, "… so that the ship should be propelled a distance of 82½ feet per minute, without considering any reduction in resistance due to an appropriate design of the prow. … When this case is adapted to a ship with a beaked prow, for which he assumed the resistance to be ten times smaller, he concluded that the speed imparted to the ship could exceed 260 feet per minute, or 15,649 feet per hour …"

These results are in frontal opposition to Euler's conclusion that the motion generated by the pendulum's impact is countered by the forces during the pendulum's ascent and subsequent descent, result in no progressive movement of the ship.

## 3. Conclusions

In comparing the approaches of Jacob Bernoulli and Euler to the problem of pendulum-induced ship propulsion, it becomes clear that Euler's method offers a more rigorous and accurate analysis. Jacob Bernoulli's approach, while innovative for its time, relies heavily on geometric interpretations and simplifies the problem by assuming that the pendulum's impulsive force is proportional to the square of its velocity, which leads to an overestimation of the force involved.

Euler, on the other hand, employs a more sophisticated approach using differential equations that better reflect the physical realities of the system. By considering the elastic properties of the platform and the continuous nature of the force exerted during the pendulum's collision with the platform, Euler provides a more accurate depiction of the forces at play. His calculation shows that the force is proportional to the velocity (or the square root of the height of fall), leading to a more realistic understanding of the forces involved.

The critical difference between the two analyses lies in how they relate the impulsive force to the fall distance $a$. Jacob Bernoulli's overestimation of the force by a factor of $\sqrt{a}$ demonstrates the limitations of his method, particularly in the context of more complex, real-world applications. Additionally, the comparison between Euler's impact force calculation and the result obtained using the Dirac Delta function suggests that Euler's inclusion of elasticity was not fully integrated into the model, highlighting a potential area for refinement in his analysis.

Ultimately, Euler's conclusion—that the forces generated by the pendulum's impact are countered by opposing forces during its ascent and subsequent descent, resulting in no progressive movement of the ship—stands in stark contrast to Jacob Bernoulli's overly optimistic projections. This comparison underscores the importance of using a comprehensive and mechanically sound approach when analyzing such problems, as demonstrated by Euler's more accurate and realistic results.

Finally, Euler provides a detailed proof that no motion can be generated by internal forces alone. Through rigorous mathematical analysis, he demonstrates that any internal mechanism within a closed system, such as a ship, cannot produce net propulsion. Euler systematically examines the interactions of forces within the system, showing that for every force exerted in one direction, there is an equal and opposite force that cancels it out. This meticulous approach leads to the conclusion that, regardless of the internal dynamics or the mechanical design, the system as a whole cannot achieve forward movement without the influence of an external force. Euler's proof effectively dispels the notion of perpetual motion or self-propelling devices, reinforcing the fundamental principles of Newtonian mechanics.

## Note

This manuscript was prepared with the assistance of the GPT-4 language model, an advanced AI developed by OpenAI. The AI was utilized to aid in drafting, reviewing, and refining the content presented herein. While every effort has been made to ensure the accuracy and clarity of the material, the final responsibility for the content rests with the author.

## Declarations

No funding was received for conducting this study.

The author has no competing interests to declare that are relevant to the content of this article.